\documentclass[letterpaper]{article}

\usepackage[T1]{fontenc}
\usepackage[utf8]{inputenc}

\usepackage{geometry}
\usepackage{setspace}
\usepackage[numbers,sort&compress]{natbib}
\usepackage{graphicx}
\usepackage{xcolor}
\usepackage{soul}

\definecolor{revblue}{RGB}{160,210,255}
\sethlcolor{revblue}
\DeclareRobustCommand{\rev}[1]{#1}

\usepackage{float}
\usepackage{amsmath}
\usepackage{amssymb}

\usepackage{chemformula}
\usepackage[version=4]{mhchem}

\usepackage{authblk}

\author[1]{G. Lee}
\author[1]{A. Borel}
\author[2]{T. Taniguchi}
\author[3]{K. Watanabe}
\author[1]{F. Sirotti}
\author[1]{F. Cadiz}

\affil[1]{Laboratoire de Physique de la Matière Condensée, CNRS,
Ecole polytechnique, Institut Polytechnique de Paris, 91120 Palaiseau, France}
\affil[2]{Research Center for Materials Nanoarchitectonics, National Institute for Materials Science, 1-1 Namiki, Tsukuba 305-0044, Japan}
\affil[3]{Research Center for Electronic and Optical Materials, National Institute for Materials Science, 1-1 Namiki, Tsukuba 305-0044, Japan}

\title{Optical Thermometry in Monolayer \ce{WS2}: Raman and
Photoluminescence Spectroscopy from 5~K to 1273~K}

\date{*Email: fabian.cadiz@polytechnique.edu}

\begin{document}

\maketitle

%%%%%%%%%%%%%%%%%%%%%%%%%%%%%%%%%%%%%%%%%%%%%%%%%%%%%%%%%%%%%%%%%%%%%
\begin{abstract}
%%%%%%%%%%%%%%%%%%%%%%%%%%%%%%%%%%%%%%%%%%%%%%%%%%%%%%%%%%%%%%%%%%%%%

Reliable temperature measurement is essential for
characterizing and engineering two-dimensional (2D) semiconductor devices, yet
optical thermometry of 2D materials has so far been demonstrated only over
narrow temperature windows spanning a few hundred kelvin. Here, we establish a comprehensive optical
thermometry framework for monolayer \ce{WS2} from 5~K to 1273~K. We show that: (i) the out-of-plane
A$_{1g}$ Raman mode shifts linearly with temperature above 100~K, with a slope of $-0.0138~\mathrm{cm^{-1}/K}$ in non-encapsulated \ce{WS2}
(fitted over 298--1223~K) and a measurably smaller slope of
$-0.0121~\mathrm{cm^{-1}/K}$ in an hBN-encapsulated sample (fitted over 100--1273~K);
(ii) photoluminescence spectra measured over the same 5--1273~K range provide
an independent optical indicator of the temperature: the neutral-exciton emission $X_A^0$
redshifts continuously with temperature and is well described by the
semi-empirical Pässler model. The two indicators probe distinct physical
mechanisms, lattice anharmonicity and bandgap renormalization, and together
provide contact-free thermometry for monolayer \ce{WS2} from 5~K to 1273~K,
an approach applicable to 2D semiconductors more broadly.

\end{abstract}

%%%%%%%%%%%%%%%%%%%%%%%%%%%%%%%%%%%%%%%%%%%%%%%%%%%%%%%%%%%%%%%%%%%%%
\section{Introduction}
%%%%%%%%%%%%%%%%%%%%%%%%%%%%%%%%%%%%%%%%%%%%%%%%%%%%%%%%%%%%%%%%%%%%%

The isolation of two-dimensional (2D) atomic crystals, beginning with graphene
and extending to a broad family of layered compounds \cite{novoselov-2005}, has
established a class of materials whose \rev{unique} properties are set by their reduced
dimensionality. Among them, monolayer transition metal dichalcogenides (TMDs)
are direct-gap semiconductors whose optical response is dominated by tightly
bound excitons even at room temperature \cite{mak-2010,splendiani-2010}. The
strong Coulomb interaction and reduced dielectric screening in these
atomically-thin materials give exciton binding energies of several hundred meV
and make the optical response acutely sensitive to the dielectric surroundings
\cite{ugeda-2014,chernikov-2014}.

High-temperature annealing \rev{of monolayers has been shown to be a promising} route
to defect engineering in 2D semiconductors: in monolayer \ce{MoS2} it generates chalcogen-vacancy defects
with characteristic optical signatures \cite{mitterreiter-2021}, and in
hBN-encapsulated monolayer \ce{WS2} annealing near 1100~K produces ultra-narrow
emission lines that act as single-photon emitters \cite{lee-2026d}. Controlling
such processes requires reliable knowledge of the temperature of the 2D layer
itself, which cannot be obtained by attaching a contact sensor to a layer which
is only a few atoms thick. The optical response of the monolayer offers a way out: both its
excitonic emission and its Raman spectrum shift measurably with temperature, so
the spectrum itself can serve as a built-in, contact-free thermometer.

Optical thermometry based on photoluminescence (PL) and Raman spectroscopy is
well established as a non-contact probe at the nanoscale \cite{brites-2012},
and has been extended to 2D materials such as graphene, where the G and 2D
Raman modes provide local temperature readout \cite{calizo-2007}. In the case of
monolayer \ce{MoS2} and \ce{WS2}, temperature-dependent Raman spectroscopy has
been used to extract thermal conductivities \cite{yan-2014,sahoo-2013}, with
linear peak shifts reported over limited temperature windows, typically
spanning a few hundred kelvin below $\sim$550~K \cite{huang-2016,lanzillo-2013}. However, the combination of (i) an ultra-wide
temperature range from cryogenic to near-dissociation temperatures, (ii)
simultaneous access to both Raman and PL observables, and (iii) a quantitative
assessment of the robustness of these indicators across different encapsulation
conditions has not been demonstrated to date.

A key practical challenge is that high-power laser illumination, required for
sufficient Raman signal in monolayer TMDs, induces strong photodoping in
non-encapsulated samples, as established for the related molybdenum
dichalcogenides \cite{cadiz-2016}, making \rev{neutral exciton} PL inaccessible after Raman
acquisition. Encapsulation of the TMD monolayer between hexagonal boron nitride
(hBN) layers suppresses this photodoping effect \cite{cadiz-2016} and
dramatically narrows excitonic linewidths \cite{cadiz-2017,ajayi-2017}, enabling PL
spectroscopy under the same conditions as Raman.

In this work, we establish an optical thermometry
framework for monolayer \ce{WS2} spanning 5~K to 1273~K---a range that, to our
knowledge, has not been covered by a single 2D material system in a single
study. Both samples are mounted on a suspended SiC micro-heater membrane inside
a cryostat, so that the monolayer can be heated locally. The refractory character of SiC allows the membrane to
be operated repeatedly up to 1273~K, so that the same sample can be measured
over the full range without changing the optical setup. The non-encapsulated sample gives the Raman thermal response from
room temperature to 1223~K, while the hBN-encapsulated sample provides PL over
the full range together with complementary Raman data. Calibrating both
signatures against a known temperature turns the spectrum itself into the
thermometer: changes in the Raman or PL signatures then report the temperature of
the monolayer at that spot.

%%%%%%%%%%%%%%%%%%%%%%%%%%%%%%%%%%%%%%%%%%%%%%%%%%%%%%%%%%%%%%%%%%%%%
\section{Experimental Methods}
%%%%%%%%%%%%%%%%%%%%%%%%%%%%%%%%%%%%%%%%%%%%%%%%%%%%%%%%%%%%%%%%%%%%%

\subsection{Sample Preparation}

For the non-encapsulated sample (Experiment 1), monolayer \ce{WS2} was
mechanically exfoliated from a bulk crystal and transferred directly onto the
suspended SiC micro-membrane of a commercial micro-heater chip (EDEN
Instruments). For the encapsulated sample (Experiment 2), hBN/\ce{WS2}/hBN
heterostructures were assembled by dry-stamp transfer \cite{castellanos-gomez-2014}
and subsequently placed onto a second membrane chip. The SiC membrane is
120~nm thick, suspended between doped Si contacts, and enables uniform Joule
heating of the central region up to $\sim$1200~K when a dc current is applied.
The temperature of the membrane was calibrated against the applied electrical
current by the manufacturer \cite{lee-2026d}.

\subsection{Temperature Control}

Two complementary temperature stages were used for the encapsulated sample.
From 5~K to room temperature, the sample was cooled in a He-flow cryostat and
heated uniformly using the cryostat's built-in heater. In this regime, \rev{a
calibrated Cernox thermometer measures the temperature of the sample holder},
which is in equilibrium with the sample. From room temperature to $\sim$1273~K, the
SiC membrane was heated resistively under vacuum ($\sim$$10^{-3}$~mbar), \rev{with
the cryostat cooling switched off so that the chip surroundings remained at room
temperature, as required for the manufacturer's calibration}. The current-to-temperature
conversion was obtained from the manufacturer's calibration. For the
non-encapsulated sample, \rev{only Raman measurements were performed, because of
photodoping effects, from room temperature to 1223~K under vacuum}.

\subsection{Optical Measurements}

All measurements used a 514.5~nm continuous-wave laser (Cobolt). For Raman
spectroscopy, the excitation power was $\sim$1~mW; for PL spectroscopy of the
encapsulated sample, the power was reduced to $\lesssim$10~$\mu$W to avoid
photo-induced heating. The laser was focused to a $\sim$1~$\mu$m spot through \rev{an
apochromatic objective with a numerical aperture of 0.82}. The scattered or emitted light was
collected by the same objective, filtered with volume Bragg notch filters centered
at 514.5~nm (for Raman), and dispersed by a grating spectrometer onto a
Peltier-cooled CCD. \rev{Starting from room temperature, the heater current was
increased gradually from one temperature step to the next, and Raman or PL
spectra were acquired in situ at each step.} PL spectra were fitted with a \rev{Voigt}
profile (see Supporting Information) to extract the peak energy $E_0$ and the
total full-width at half-maximum (FWHM) at each temperature.

%%%%%%%%%%%%%%%%%%%%%%%%%%%%%%%%%%%%%%%%%%%%%%%%%%%%%%%%%%%%%%%%%%%%%
\section{Results}
%%%%%%%%%%%%%%%%%%%%%%%%%%%%%%%%%%%%%%%%%%%%%%%%%%%%%%%%%%%%%%%%%%%%%

\subsection{Raman Spectroscopy as a Thermal Indicator}

\subsubsection{Non-encapsulated \ce{WS2} (RT to 1223~K)}

\begin{figure}[H]
  \centering
  \includegraphics[width=\linewidth]{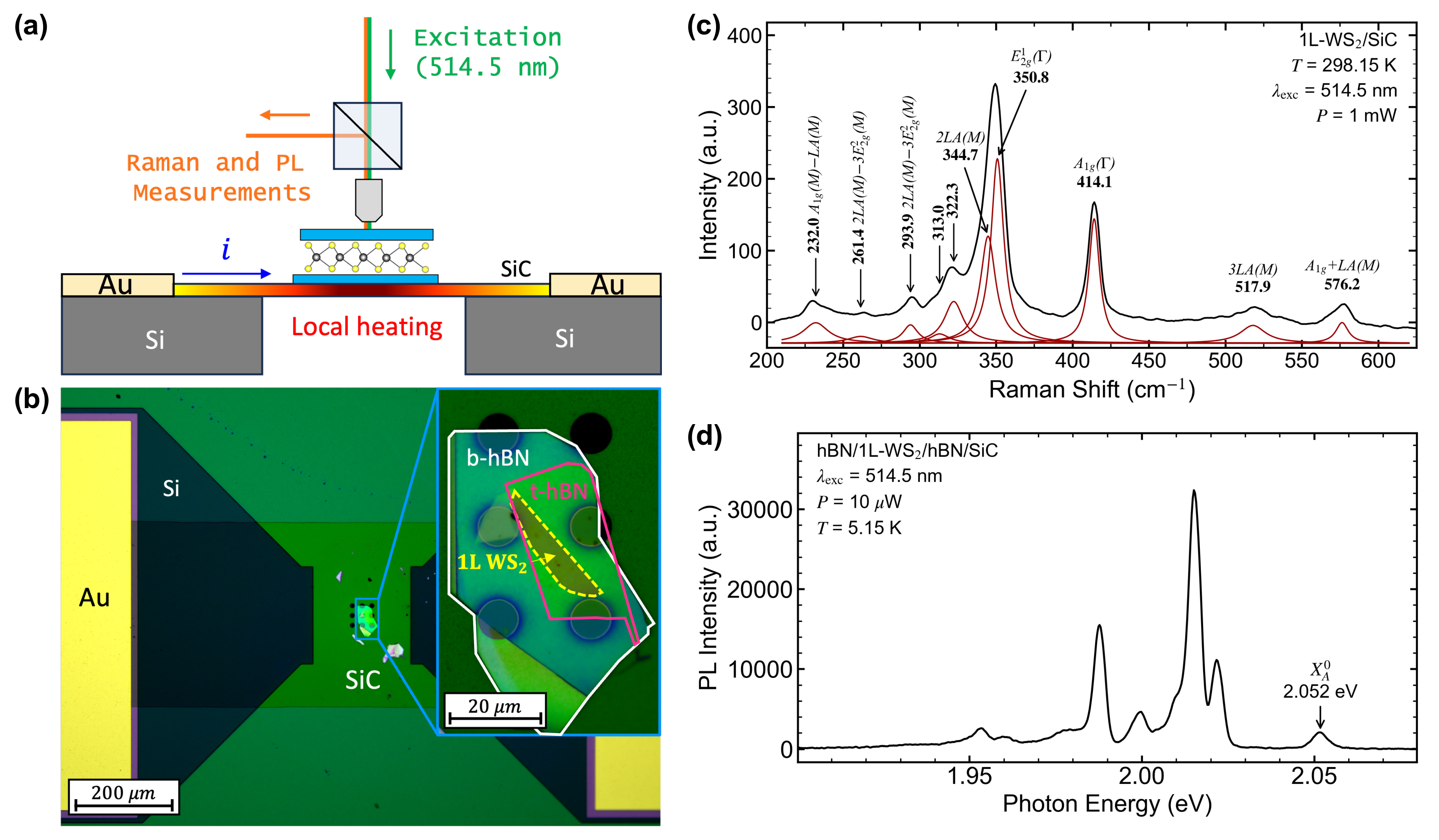}
  \caption{\textbf{Experimental platform and reference spectra.}
    (a) Schematic of the in-situ measurement setup: hBN/\ce{WS2}/hBN
    heterostructure transferred onto the suspended SiC micro-membrane of the
    micro-heater chip, mounted inside a cryostat.
    (b) Optical image of the encapsulated \ce{WS2} flake on the SiC membrane
    under white-light illumination.
    (c) Room-temperature Raman spectrum of non-encapsulated monolayer \ce{WS2}
    ($\lambda_\mathrm{exc} = 514.5$~nm, $P = 1$~mW), including Lorentzian peak
    fits for the E$^1_{2g}$($\Gamma$), 2LA(M), A$_{1g}$($\Gamma$), and
    higher-order combination modes.
    High-power illumination induces photodoping in the absence of hBN encapsulation,
    precluding PL measurements on this sample.
    (d) PL spectrum of hBN-encapsulated \ce{WS2} at 5~K, showing the neutral
    exciton $X_A^0$, trion, and dark-exciton-related features
    ($\lambda_\mathrm{ex} = 514.5$~nm, $P = 10$~$\mu$W).}
  \label{fig:fig1}
\end{figure}

A representative Raman spectrum of the non-encapsulated \ce{WS2} monolayer at
room temperature is shown in Fig.~\ref{fig:fig1}(c). In addition to the
principal zone-center modes E$^1_{2g}$($\Gamma$) ($\sim$352~cm$^{-1}$) and
A$_{1g}$($\Gamma$) ($\sim$414~cm$^{-1}$), several zone-boundary combination
modes are resolved, including 2LA(M) ($\sim$349~cm$^{-1}$), 3LA(M)
($\sim$517~cm$^{-1}$), and A$_{1g}$($\Gamma$)+LA(M) ($\sim$576~cm$^{-1}$),
consistent with prior reports on monolayer \ce{WS2} \cite{berkdemir-2013}.
The A$_{1g}$($\Gamma$) mode is used as the primary thermal indicator in what
follows; the overlapping E$^1_{2g}$($\Gamma$)+2LA(M) doublet is analysed
separately in the Supporting Information (Sec.~S1), where it serves as a
secondary, qualitative check.

Figure~\ref{fig:fig2}(a) shows a two-dimensional intensity map of the Raman
spectra as a function of temperature from room temperature to 1223~K.
With increasing temperature, both peaks redshift continuously and broaden,
while their intensity decreases. The redshift arises from two contributions:
thermal expansion of the lattice, which lowers the phonon frequency through the
Grüneisen relation, and anharmonic phonon--phonon coupling, whose three- and
four-phonon decay channels become increasingly populated \cite{huang-2016}. The signal vanishes above
$\sim$1223~K, consistent with thermal dissociation of the
\ce{WS2} lattice through sulfur sublimation \cite{brainard-1968} (see also
Supporting Information).

The A$_{1g}$ peak position extracted from these spectra is shown in
Fig.~\ref{fig:fig2}(c) (red squares). A linear fit restricted to $T \geq 100$~K
yields:

\begin{equation}
\omega_{A_{1g}}(T) = \omega_0 + \left(\frac{d\omega}{dT}\right) T,
\end{equation}

with a slope $d\omega/dT = -0.0138 \pm 0.0002~\mathrm{cm^{-1}/K}$ and intercept
$\omega_0 = 417.95~\mathrm{cm^{-1}}$ ($R^2 = 0.995$, $T = 298$--$1223$~K); the
same dataset analysed with the anti-Stokes A$_{1g}$ line included gives
$-0.0134~\mathrm{cm^{-1}/K}$ \cite{lee-2026d}. A linear fit to the
A$_{1g}$ shift of monolayer \ce{WS2} on quartz by Huang \textit{et al.} gave
$-0.0122~\mathrm{cm^{-1}/K}$ over 84--543~K, although those authors restricted
their quantitative analysis of this mode to below 380~K, above which strain
from wrinkling of the monolayer on the substrate distorted the response
\cite{huang-2016}. Reported first-order coefficients for the A$_{1g}$ mode
span $-0.006$ to $-0.0149~\mathrm{cm^{-1}/K}$ across studies \cite{huang-2016},
so that values obtained on different samples, substrates and fitting ranges are
not directly comparable; the two configurations measured here were acquired on
the same platform with identical excitation and fitting procedures, making the
difference between them, quantified below, a controlled comparison. The A$_{1g}$ mode is preferred as the
primary thermometric indicator because, unlike the overlapping
E$^1_{2g}$($\Gamma$)+2LA(M) doublet, it is spectrally isolated, so that its
position can be determined unambiguously at every temperature.

\subsubsection{hBN-encapsulated \ce{WS2} (5~K to 1273~K)}

The hBN-encapsulated sample allows simultaneous Raman and PL measurements under
identical laser conditions, since hBN encapsulation suppresses the photodoping
that precludes \rev{neutral exciton} PL in the non-encapsulated sample \cite{cadiz-2016}.
The Raman and PL data presented in this and the following section were acquired
on the same encapsulated sample.

\begin{figure}[H]
  \centering
  \includegraphics[width=\linewidth]{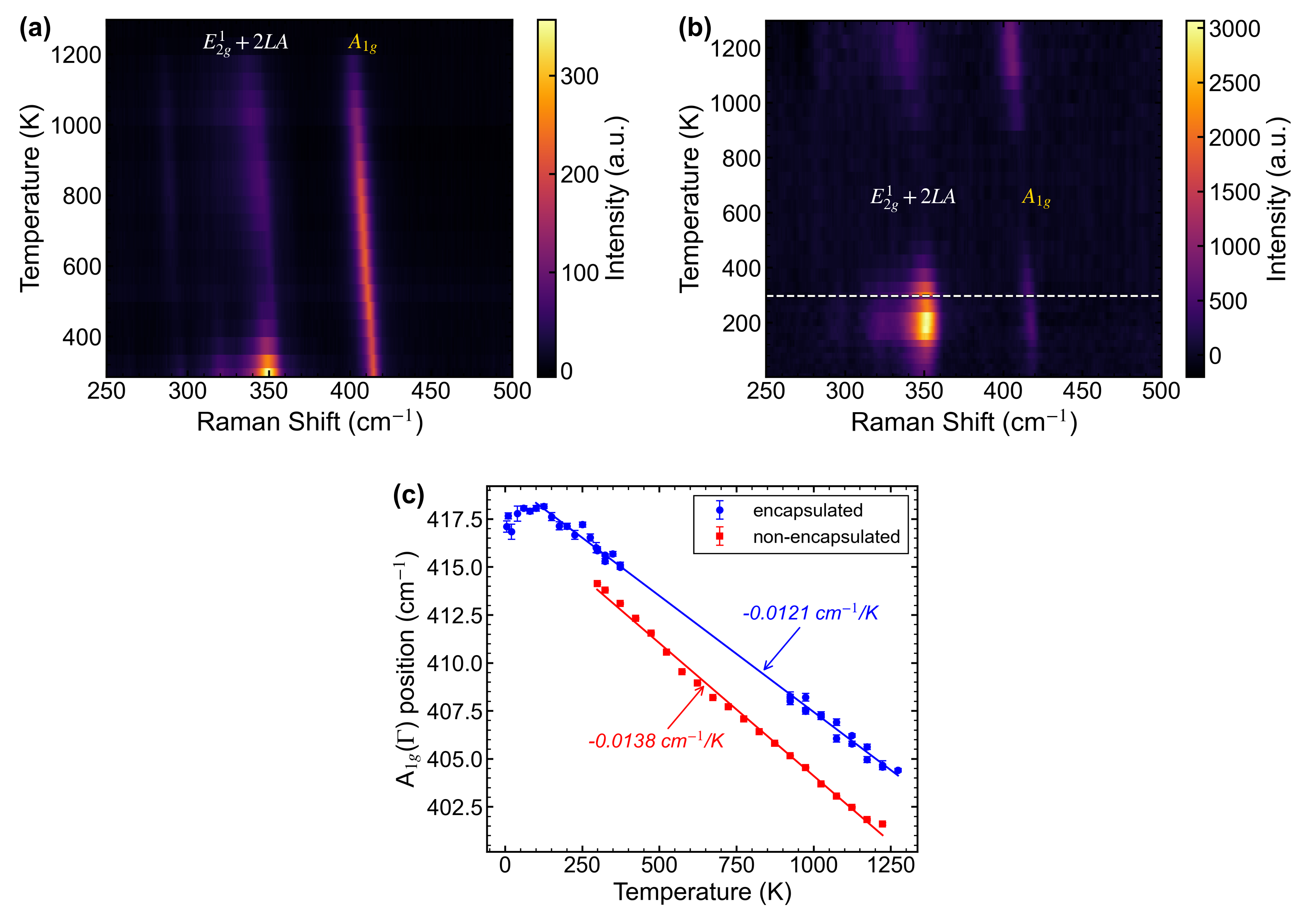}
  \caption{\textbf{Raman thermal indicator: non-encapsulated and hBN-encapsulated
    \ce{WS2}.}
    (a) Two-dimensional map of Raman intensity versus temperature (298--1223~K)
    and Raman shift for the non-encapsulated sample. Both the E$^1_{2g}$+2LA and
    A$_{1g}$ modes redshift continuously with increasing temperature.
    (b) Two-dimensional map of Raman intensity versus temperature (5--1273~K)
    and Raman shift for the hBN-encapsulated sample. The white dashed line marks
    the boundary between the cryostat (5~K--RT) and micro-heater (RT--1273~K)
    acquisition stages. A temperature-dependent intensity redistribution between
    the E$^1_{2g}$+2LA (dominant at low $T$) and A$_{1g}$ (dominant at high $T$)
    modes is observed, with both peaks weak at intermediate temperatures.
    (c) A$_{1g}$($\Gamma$) peak position versus temperature extracted from both
    samples: non-encapsulated (red squares) and hBN-encapsulated (blue circles).
    Solid lines are linear fits restricted to $T \geq 100$~K, with slopes of
    $-0.0138~\mathrm{cm^{-1}/K}$ (non-encapsulated, $R^2 = 0.995$) and
    $-0.0121~\mathrm{cm^{-1}/K}$ (hBN-encapsulated, $R^2 = 0.997$). Raman and PL
    measurements on the encapsulated sample were performed simultaneously (see
    Fig.~\ref{fig:fig3}).}
  \label{fig:fig2}
\end{figure}

Figure~\ref{fig:fig2}(b) shows the two-dimensional Raman intensity map of the
encapsulated sample over the full temperature range from 5~K to
1273~K. \rev{The overall Raman signal is weaker than for the non-encapsulated
sample. With a single sample of each type, we do not attribute this difference
to encapsulation itself: the hBN/}\ce{WS2}\rev{/hBN stack on the thin SiC membrane
forms a multilayer optical cavity in which interference of both the excitation
and the scattered light depends on the layer thicknesses, so that the Raman
signal can be either reduced or enhanced depending on the stack geometry.} The map reveals a
temperature-dependent intensity redistribution between the two modes: at low
temperature (below \rev{$\sim$400~K}), the E$^1_{2g}$+2LA mode dominates; at high
temperature, the A$_{1g}$ mode becomes the stronger feature; at intermediate
temperatures both modes are weak, leading to gaps in the extracted peak
positions. \rev{This redistribution is consistent with a change in resonance
condition. The excitation energy (2.410~eV) lies close to the B exciton of
monolayer} \ce{WS2}\rev{, and Raman scattering in} \ce{WS2} \rev{is enhanced near
the excitonic resonances} \cite{delcorro-2016}\rev{, most strongly for the
double-resonant 2LA(M) band} \cite{berkdemir-2013}. \rev{As the excitonic
transitions redshift with temperature, the E$^1_{2g}$+2LA intensity of the
encapsulated sample peaks near 225~K, where the B exciton crosses the
excitation energy, and both modes weaken as the detuning grows. The partial
recovery above $\sim$900~K may reflect the approach of the higher-lying C
exciton. This regime is not reached in the non-encapsulated sample, whose
Raman signal vanishes above $\sim$1223~K as the monolayer dissociates
(Supporting Information, Sec.~S6), whereas the encapsulated monolayer
survives to 1273~K. A detailed analysis is given in the Supporting
Information (Sec.~S7).}

Despite the missing data points at intermediate temperatures,
Fig.~\ref{fig:fig2}(c) (blue circles) shows that the A$_{1g}$ peak positions
obtained from the encapsulated sample across the full accessible range (5~K to
RT from cryogenic measurements; RT to 1273~K from micro-heater measurements)
follow the same qualitative linear trend established for the non-encapsulated
sample, indicating that the underlying phonon-softening mechanism is intrinsic
to the \ce{WS2} lattice regardless of encapsulation, although the two samples
differ measurably in thermal sensitivity, as quantified below. A linear fit
restricted to $T \geq 100$~K yields a slope
$d\omega/dT = -0.0121 \pm 0.0001~\mathrm{cm^{-1}/K}$ and intercept
$\omega_0 = 419.55~\mathrm{cm^{-1}}$ ($R^2 = 0.997$), a magnitude
$\sim$12\% smaller than in the non-encapsulated sample. Because this fit pools
cryostat-calibrated points below room temperature with micro-heater-calibrated
points above it, whereas the non-encapsulated fit uses micro-heater
calibration alone, we checked that the difference is not an artefact of
combining the two temperature scales: restricting the encapsulated fit to
$T \geq 298$~K, so that it too rests on the micro-heater calibration only,
gives $-0.0121~\mathrm{cm^{-1}/K}$, unchanged within the fit uncertainty.
This difference is in
the direction expected if the hBN layers mechanically constrain the monolayer:
the in-plane thermal expansion coefficient of hBN is negative
\cite{androulidakis-2020}, whereas that of \ce{WS2} is positive
\cite{huang-2016}, so an encapsulated monolayer cannot expand as freely as a
bare one and the thermal-expansion contribution to the phonon softening is
correspondingly reduced. Testing this quantitatively would require an
independent measurement of the residual strain and is left for future work.

\subsection{Photoluminescence as a Thermal Indicator}

\begin{figure}[H]
  \centering
  \includegraphics[width=\linewidth]{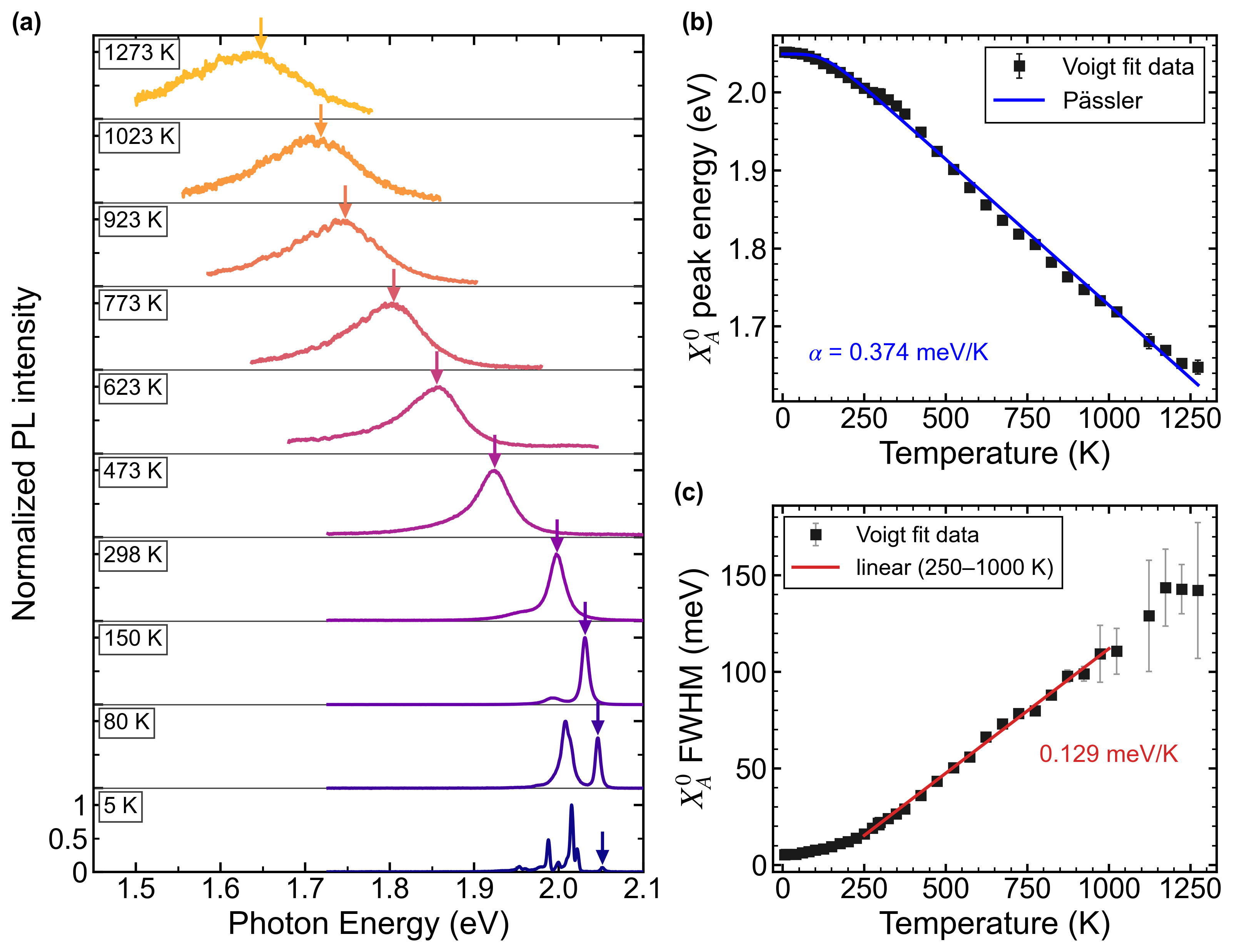}
  \caption{\textbf{Photoluminescence thermal indicators in hBN-encapsulated monolayer
    \ce{WS2}.}
    (a) Representative PL spectra from 5~K to 1273~K, each normalized to
    its own maximum. Arrows mark the $X_A^0$ energy obtained from Voigt fits;
    below $\sim$100~K the trion and localized-exciton lines on its low-energy
    side are stronger than $X_A^0$. The
    $X_A^0$ peak redshifts and
    broadens continuously with increasing temperature.
    (b) $X_A^0$ peak energy $E_0$ versus temperature from Voigt fits (black
    squares, with error bars) and the Pässler model fit (blue line),
    Eq.~\eqref{eq:passler}; $\alpha$ is the high-temperature slope. (c) Total Voigt FWHM of $X_A^0$ versus temperature (squares) and a
    linear fit over 250--1000~K (red line).}
  \label{fig:fig3}
\end{figure}

\textbf{Spectral evolution and peak energy.}
Figure~\ref{fig:fig3}(a) shows the evolution of the PL spectrum of the
hBN-encapsulated \ce{WS2} monolayer through ten representative
spectra, each normalized to its own maximum, spanning 5~K to 1273~K; all
measured spectra are shown as a two-dimensional map in Supporting Information,
Fig.~S3(a). At 5~K the emission is dominated
by lower-lying excitonic complexes, and the neutral exciton $X_A^0$ at
$E_0 = 2.052$~eV, with a narrow Voigt linewidth of $\sim$5.3~meV, is one of the
weakest features of the spectrum (Fig.~\ref{fig:fig1}(d)). This reflects the
darkish nature of tungsten-based monolayers, in which the excitonic ground
state is optically inactive, so that photocarriers relax preferentially into
dark and charged states \cite{zinkiewicz-2020,zhang-2015}. With increasing
temperature, thermal activation redistributes population into the bright
$X_A^0$ state, whose integrated intensity rises by two orders of magnitude up
to $\sim$175~K (Supporting Information, Fig.~S5); at the same time the
lower-lying complexes broaden and merge, and above $\sim$300~K the spectrum
consists of a single $X_A^0$ line. On further heating this line redshifts
monotonically and broadens, while its emission efficiency falls through
thermally activated non-radiative recombination \cite{korn-2011} and the escape
of excitons from the radiative window \cite{robert-2016}.

The neutral-exciton peak energy $E_0(T)$ extracted from Voigt fits is shown in
Fig.~\ref{fig:fig3}(b). The redshift reflects the interplay of two mechanisms:
thermal lattice expansion, which reduces the orbital overlap and lowers the
bandgap, and electron--phonon coupling, which renormalizes the quasiparticle
energies \cite{passler-1999}. The data are excellently described by the
Pässler model \cite{passler-1997,passler-1999}:

\begin{equation}
E_0(T) = E_0(0) - \frac{\alpha \Theta}{2}
\left[ \sqrt[p]{1 + \left(\frac{2T}{\Theta}\right)^p} - 1 \right],
\label{eq:passler}
\end{equation}

where $E_0(0)$ is the zero-temperature bandgap, $\alpha$ is the high-temperature
slope, $\Theta$ is the mean phonon temperature, and $p$ is a dimensionless
dispersion exponent. The best-fit parameters obtained here are:
$E_0(0) = 2.0491 \pm 0.0036$~eV, $\alpha = (3.741 \pm 0.069)\times10^{-4}~\mathrm{eV/K}$,
$\Theta = 280.1 \pm 28.8$~K, $p = 4.04 \pm 2.23$ (the large relative uncertainty
on $p$ reflects its known correlation with $\Theta$ in this model)
($R^2 = 0.9964$, RMSE $= 7.9$~meV;
see Table~S1). These are consistent with previously reported values
for encapsulated \ce{WS2} \cite{lee-2026d}, with small deviations attributed to
differences in local strain and dielectric environment between samples. A
comparison of the Pässler model with the Varshni and O'Donnell--Chen models is
provided in the Supporting Information; both the Pässler and O'Donnell--Chen
models reproduce the data equivalently well, while the Varshni model
underestimates the curvature at low temperature.

The inversion of Eq.~\eqref{eq:passler} allows the temperature to be read
directly from the measured $E_0$, providing a model-based, contact-free
thermometer calibrated against the cryostat's Cernox thermometer (5~K to RT) and the manufacturer's
chip calibration (RT to 1273~K). The practical lower limit of this PL
thermometer is set by the spectral resolution of the detection system. Towards
higher temperature, the $X_A^0$ peak broadens substantially (from
$\sim$5.3~meV at 5~K to $\sim$140~meV at 1273~K) and becomes progressively less
well defined, so that its position is determined with decreasing precision;
the measurements reported here extend to 1273~K.
The root-mean-square deviation of the Pässler fit residuals is
$7.9$~meV, corresponding to a temperature uncertainty of $\sim 21$~K in the
high-temperature (linear) regime, estimated as RMSE$/\alpha$; the uncertainty
is larger at cryogenic temperature, where $dE_0/dT \to 0$.

\textbf{Spectral linewidth.}
The $X_A^0$ emission broadens monotonically with temperature, the total Voigt
FWHM increasing from $\sim$5.3~meV at 5~K to $\sim$140~meV at 1273~K
(Fig.~\ref{fig:fig3}(c)). The narrow low-temperature linewidth
reflects the quality of the hBN-encapsulated heterostructure
\cite{cadiz-2017,ajayi-2017}, while the progressive broadening at higher
temperature is consistent with an increasing phonon-scattering-limited
dephasing rate \cite{selig-2016,moody-2015}; a fit of the full
5--1000~K dependence with a phonon-scattering model is given in Supporting
Information, Fig.~S5. \rev{Above $\sim$250~K the FWHM
increases linearly with temperature, with a slope of $0.129 \pm 0.002$~meV/K
($R^2 = 0.997$ over 250--1000~K; Fig.~\ref{fig:fig3}(c)). The
linewidth therefore provides an additional PL-based thermometric indicator:
the rms residual of 1.7~meV corresponds to a temperature uncertainty of
$\sim$13~K, comparable to that obtained from the peak energy. As for the peak
energy and the Raman shift, the calibration is specific to the sample. Above
$\sim$1100~K the width levels off as the low-energy tail of the peak extends
beyond the fitting window, which sets the upper limit of this indicator.}

%%%%%%%%%%%%%%%%%%%%%%%%%%%%%%%%%%%%%%%%%%%%%%%%%%%%%%%%%%%%%%%%%%%%%
\section{Discussion}
%%%%%%%%%%%%%%%%%%%%%%%%%%%%%%%%%%%%%%%%%%%%%%%%%%%%%%%%%%%%%%%%%%%%%

Our results establish monolayer \ce{WS2} as a broadband optical thermometer
whose two independent indicators---Raman peak position and PL peak energy---are
applicable over different, overlapping temperature windows that together span
5~K to 1273~K.

The Raman indicator is the simpler of the two: a single linear calibration
curve suffices above 100~K. The hBN-encapsulated sample, however, shows a
measurably different calibration from the bare monolayer, its slope being
12\% smaller ($-0.0121$ vs.\ $-0.0138~\mathrm{cm^{-1}/K}$). Because the slope
is a differential quantity, this difference is independent of the absolute
wavenumber calibration, and it sets a floor on the accuracy of any transferred
calibration: using one sample's slope to read the other's temperature leaves a
systematic error of 12\%, or $\sim$120~K at 1000~K. A Raman calibration
obtained on a bare monolayer therefore cannot be carried over to an
encapsulated heterostructure without re-calibration. What this work establishes
is the range over which either calibration remains linear, from 100~K to
1273~K, substantially broader than the window previously reported for
\ce{WS2} Raman thermometry \cite{huang-2016}.

The PL thermometer relies on the neutral-exciton peak energy, for which the
Pässler model provides a physically grounded, analytic expression that can be
inverted to yield temperature from a single spectral acquisition.
Importantly, the PL thermometer extends the range to cryogenic temperatures,
which the Raman thermometer cannot easily access due to the difficulty of
distinguishing small thermal shifts in the nearly temperature-independent phonon
spectrum at low $T$.

The two indicators are complementary in a specific sense. \rev{In terms of
sensitivity, the PL peak energy is by far the more responsive: in the linear
regime it shifts by $\alpha \approx 0.37$~meV/K, compared with
$|d\omega/dT| = 0.0121$~cm$^{-1}$/K ($\approx 1.5~\mu$eV/K) for the A$_{1g}$
mode, corresponding to relative sensitivities $(1/X)|dX/dT|$ of
$\approx 2\times10^{-4}$~K$^{-1}$ and $\approx 3\times10^{-5}$~K$^{-1}$,
respectively. Between $\sim$80~K and room temperature, where the $X_A^0$ line
is narrow, this yields a statistical temperature uncertainty below 1~K,}
whereas the Raman shift is too weakly temperature-dependent to resolve below
$\sim$100~K. \rev{Below $\sim$40~K, $dE_0/dT$ itself vanishes and the PL
resolution degrades to several kelvin and more. At high temperature, the
advantage in sensitivity is offset by the broadening of the PL line, which
increases the uncertainty of the fitted peak position:} converting the
peak-position uncertainty of each fitted spectrum
into a temperature through the corresponding sensitivity ($\sigma_{E_0}/\alpha$
for PL, $\sigma_\omega/|d\omega/dT|$ for Raman) gives, averaged over all points
above 900~K, $\sim$12~K for PL and $\sim$13~K for Raman. The practical
distinction there is
not precision but sample configuration, since photodoping precludes \rev{neutral exciton} PL on
non-encapsulated samples \cite{cadiz-2016}, leaving the Raman shift as the only
available indicator for bare monolayers.

Finally, it is useful to place this range in the context of Raman thermometry
on other 2D materials, where the published temperature windows are
considerably narrower: 100--320~K for suspended monolayer \ce{MoS2}
\cite{yan-2014}, 303--448~K for monolayer \ce{MoS2} on \ce{SiO2}
\cite{lanzillo-2013}, and 84--543~K for monolayer \ce{WS2} on quartz
\cite{huang-2016}, the last restricted to below 380~K for the A$_{1g}$ mode by
substrate-induced wrinkling. The linear behaviour reported here therefore
extends over a range several times wider than in any of these studies.
Departures from linearity are present but small: including a quadratic term
changes the fitted frequency by at most 0.4~cm$^{-1}$ out of a total excursion
of 12.5~cm$^{-1}$, so a single linear calibration is retained throughout.
\rev{Other optical signatures of the same material can offer higher sensitivity
over narrower windows: anti-Stokes (upconversion) PL in monolayer} \ce{WS2}
\rev{reaches a relative sensitivity above 4\%~K$^{-1}$ between 300 and 425~K and
has been used to map a thermal gradient across a single flake with $\sim$1~$\mu$m
spatial resolution} \cite{nagarkar-2026}. \rev{The Stokes PL and Raman calibrations
reported here are less sensitive but cover the full 5--1273~K range.}

The two samples reach their upper limits for different reasons. The
non-encapsulated monolayer dissociates through sulfur sublimation
\cite{brainard-1968}: its Raman signal is lost above $\sim$1223~K and the flake
is no longer visible afterwards (Supporting Information, Sec.~S6). The
encapsulated monolayer survives 1273~K intact, with both Raman and PL still
measurable, and the measurement was stopped there to protect the heater chip
rather than because the sample failed. hBN encapsulation therefore extends the
thermal stability of the monolayer as well as its spectral quality.

Two features of the platform make the extended range accessible: the suspended SiC membrane allows the monolayer to
be heated locally without defocusing or producing a spatial shift of the
flake under the laser spot, and hBN
encapsulation sharpens the excitonic linewidth sufficiently for reliable peak
fitting down to 5~K \cite{cadiz-2017,ajayi-2017}. The large exciton binding
energy of monolayer \ce{WS2} ($\sim$300~meV) \cite{chernikov-2014} ensures a
well-defined excitonic peak at all temperatures studied, and the use of a
single excitation wavelength (514.5~nm) for both Raman and PL simplifies the
measurement.

%%%%%%%%%%%%%%%%%%%%%%%%%%%%%%%%%%%%%%%%%%%%%%%%%%%%%%%%%%%%%%%%%%%%%
\section{Conclusion}
%%%%%%%%%%%%%%%%%%%%%%%%%%%%%%%%%%%%%%%%%%%%%%%%%%%%%%%%%%%%%%%%%%%%%

We have established an optical thermometry framework for monolayer \ce{WS2}
spanning 5~K to 1273~K using two independent optical indicators. The
A$_{1g}$ Raman mode shifts linearly with temperature above 100~K, with slopes
of $-0.0138~\mathrm{cm^{-1}/K}$ for the bare monolayer (298--1223~K) and
$-0.0121~\mathrm{cm^{-1}/K}$ for the hBN-encapsulated heterostructure
(100--1273~K); the 12\% difference between them means that a calibration
obtained on one configuration cannot be transferred directly to the other. The
neutral-exciton PL peak energy, described by the Pässler model, provides a
second indicator over the same range, with a fit residual corresponding to an uncertainty of $\sim$21~K in the high-temperature regime\rev{; its linewidth, which increases
linearly from 250 to 1000~K, provides an additional indicator with a
comparable uncertainty of $\sim$13~K}.

\rev{One question remains open:} the upper limit reached here was set by the
heater chip rather than by the encapsulated monolayer, which survived 1273~K
intact\rev{, and} how much further hBN encapsulation extends the usable range remains to
be established.

%%%%%%%%%%%%%%%%%%%%%%%%%%%%%%%%%%%%%%%%%%%%%%%%%%%%%%%%%%%%%%%%%%%%%
\section*{Acknowledgements}
%%%%%%%%%%%%%%%%%%%%%%%%%%%%%%%%%%%%%%%%%%%%%%%%%%%%%%%%%%%%%%%%%%%%%

We thank Hadrien Duprez for demonstrating the sample fabrication
procedure, and Samy Annabi and Hannes Riechert for their help with wire
bonding of the samples.
FC acknowledges support from the European Research Council (ERC Starting Grant
OneSPIN, Grant No.\ 101075855). We also acknowledge financial support from
DEEP2D (ANR-22-CE09-0013). K.W. and T.T. acknowledge support from the CREST (JPMJCR24A5), JST and World Premier International Research Center Initiative (WPI), MEXT, Japan.

%%%%%%%%%%%%%%%%%%%%%%%%%%%%%%%%%%%%%%%%%%%%%%%%%%%%%%%%%%%%%%%%%%%%%
\section*{Data Availability}
%%%%%%%%%%%%%%%%%%%%%%%%%%%%%%%%%%%%%%%%%%%%%%%%%%%%%%%%%%%%%%%%%%%%%

The data that support the findings of this study is available at https://doi.org/10.5281/zenodo.23396397.

%%%%%%%%%%%%%%%%%%%%%%%%%%%%%%%%%%%%%%%%%%%%%%%%%%%%%%%%%%%%%%%%%%%%%
\section*{Supporting Information}
%%%%%%%%%%%%%%%%%%%%%%%%%%%%%%%%%%%%%%%%%%%%%%%%%%%%%%%%%%%%%%%%%%%%%

The Supporting Information contains: (S1) Individual versus merged-peak
fitting of the E$^1_{2g}$+2LA Raman doublet for both samples; (S2)
Temperature-dependent Raman spectra of the hBN-encapsulated sample; (S3)
Bandgap model comparison (Varshni, Pässler, O'Donnell--Chen); (S4) Voigt fitting procedure and representative fit quality; (S5)
Temperature dependence of the PL spectral linewidth and integrated $X_A^0$
intensity; (S6) \ce{WS2} monolayer after thermal
dissociation\rev{; (S7) Raman intensity and excitonic resonances}.

%%%%%%%%%%%%%%%%%%%%%%%%%%%%%%%%%%%%%%%%%%%%%%%%%%%%%%%%%%%%%%%%%%%%%
\bibliography{references}

%%%%%%%%%%%%%%%%%%%%%%%%%%%%%%%%%%%%%%%%%%%%%%%%%%%%%%%%%%%%%%%%%%%%%

\end{document}